\documentclass[twocolumn,showkeys,preprintnumbers]{revtex4-2}
\usepackage{amssymb}
\usepackage{amsmath}
\usepackage{graphicx}
\usepackage{epstopdf}
\usepackage{dcolumn}
\usepackage{bm}
\usepackage{color}
\usepackage{soul}

\begin{document}

\title{Nonlinear bandgap transmission by discrete rogue waves induced in a pendulum chain}
\author{ Alain B. Togueu Motcheyo$^{1,2,*}$, Masayuki Kimura$^{3}$,  Yusuke Doi$^{4}$, Juan F. R. Archilla$^{5}$}

\affiliation{$^{1}$ Department of Mechanical Engineering, Higher
Technical Teacher's Training College
(ENSET) Ebolowa, University of Ebolowa, P.O. Box 886, Ebolowa-Cameroon\\
 $^{2}$Laboratory of Mechanics, Department of Physics, Faculty of
Science, University
of Yaounde I, P.O. Box 812, Yaounde, Cameroon\\
 $^{3}$ Department of Electrical and Electronic Engineering,
Faculty of Science and Engineering, Setsunan University, Japan
\\$^{4}$ Department of Mechanical Engineering, Graduate School of
Engineering, Osaka University 2-1 Yamadaoka, Suita, Osaka
565-0871, Japan
\\$^{5}$ Group of Nonlinear Physics, Universidad de Sevilla, ETSI
Inform\'atica, Avda Reina Mercedes s/n, 41012-Sevilla, Spain}
\email[Corresponding author: ]{alain.togueu@univ-yaounde1.cm}
  \email[\\Alternate electronic addresses:\\ ]{abtogueu@yahoo.fr; alainbertrandtogueu@gmail.com}

\begin{abstract}
We study numerically a discrete, nonlinear lattice, which is
formed by a chain of pendula submitted to a harmonic-driving
source with constant amplitude and parametrical excitation. A
supratransmission phenomenon is obtained after the derivation of
the homoclinic threshold for the case when the lattice is driven
at one edge. The lattice traps gap solitons when the chain is
subjected to a periodic horizontal displacement of the pivot.
Discrete rogue waves are generated for the case when the pendulum
is simultaneously driven and shaken. This work may pave the way
for experimental generation of discrete rogue waves within simple
devices.
\end{abstract}


\begin{keywords}
{Discrete rogue wave,
 Nonlinear supratransmission, Nonlinear pendulum}
 \end{keywords}
 \maketitle

\section{\label{sec:1}Introduction}
Since the pioneering work by Geniet and L\'eon
\cite{GenietLeonPRL02} on the nonlinear supratransmission
phenomenon, the behavior of a plane wave within the forbidden band
has fascinated a number of researchers. Nowadays, there is another
important consequence of this phenomenon, which is the capacity to
reveal the types of waves that can propagate in the dispersive
nonlinear lattice with a periodically driven boundary edge.
Breathers have been generated in mechanical systems
\cite{GenietLeonPRL02,JEMaciasDiasPlA2008,Macias-DiazJE08,MaciasDiasCNSNS17,WatanabeNishidaPLA2018,Macias-Diaztogueu20,BountisEPJSTop},
in the Fermi-Pasta-Ulam Tsingou (FPUT) model
\cite{KhomerikiLepriRuffoPRE04,AlimaMarquieBodo2017,NendjiTogueuPoP22},
 in  a discrete inductance-capacitance electrical line
\cite{TsemarquiePRE07}, in Josephson junctions \cite{Santis2022},
and in molecular dynamics models \cite{Zakharov}. An envelope
soliton has been created in
 optical wave\-guide arrays
\cite{KhomerikiPRL04,TogueuGAPcncns17,Susanto08} and in electrical
lattices
\cite{TogueuTchawouaSieweCommunNonlinearSci13,TogueuTchawouaPRE13}.
A train of dark solitons  has been generated  in a discrete
Schr\"odinger  lattice with  cubic-quintic nonlinearity
\cite{TogueuKimuraNODY19}. A travelling asymmetric bright
soliton has  been  generated for the $\alpha$, $\beta$-FPUT
\cite{NendjiTogueuPoP22}. Up to now, to the best of our
knowledge, the transmission of gap solitons is done continuously
in time due to the periodic excitation at the edge of the lattice.
There exist waves that seem to appear out of nowhere and then
vanish without a trace \cite{AkhmedievPLA2009}. They are called rogue
waves, a phenomenon that has been observed in water
\cite{ChabchoubPRL2011,ChabchoubPRX2012}, nonlinear optics
\cite{TlidiAOP2022}, photonic lattices \cite{RivasSR2020},
metamaterials \cite{OnanaPRE2014}, and in beam--plasma interactions
\cite{GPVeldesJO2013}, to mention a few systems.

Generally, rogue waves occur within systems modeled by an
analytical integrable equation.  For the continuous integrable
equations, we can name the Sine-Gordon
equation\,\cite{HouRJPhys2020}, the one-dimensional Nonlinear
Schr\"odinger (NLS) equation
\cite{AkhmedievPLA2009,ChabchoubPRL2011}, the coupled NLS equation
\cite{BaronioPRL2012}, the Coupled Higgs Equation
\cite{MuQinJPSJ2012}, and the Sasa-Satsuma equation
\cite{AkhmedievPhysicaD2015}. For the analytically integrable
discrete equation, rogue waves have been found in the discrete
Ablowitz-Ladik \cite{AkhmedievPRE2010}, the coupled Ablowitz-Ladik
equations \cite{YongMalomedCHAOS2016},  and the
Ablowitz-Musslimani equation \cite{YuCHAOS2017}. For the
non-analytical integrable discrete equation, rogue waves have been
simulated numerically in the discrete NLS equations with cubic
\cite{EfeYucePLA2015}, and saturable  \cite{TchinangTogueuPLA}
nonlinearities. We have recently pointed out the first idea for
the creation of a rogue wave within a nonlinear band gap in the
conference paper \cite{TogueuKimuraDoiJuanNOLTA2022}. The purpose
of this letter is to consider a way of creating discrete rogue
waves using a relatively simple device.

The outline of the paper is the following: in  Section
\ref{sec:2}, we present the model under investigation. The
homoclinic supratransmission threshold is derived for the discrete
Sine-Gordon equation. In Section \ref{sec:3}, we numerically
integrate the dimensionless equation governing the physical model.
Firstly, the chain of discrete pendula is studied without the
parametric excitation in order to validate the supratransmission
threshold; secondly, the lattice is shaken without a driven edge
and thirdly, the behavior of the lattice is observed with a shaken
and a driven edge simultaneously. The spectral analysis for the
latter system is also included. Section \ref{sec:4} concludes the
letter.
\newpage
\section{Model and supratransmission threshold}\label{sec:2}
\subsection{Model description}\label{subsec:2}
\begin{figure}[ht]
\begin{center}
\includegraphics[width=3.0in]{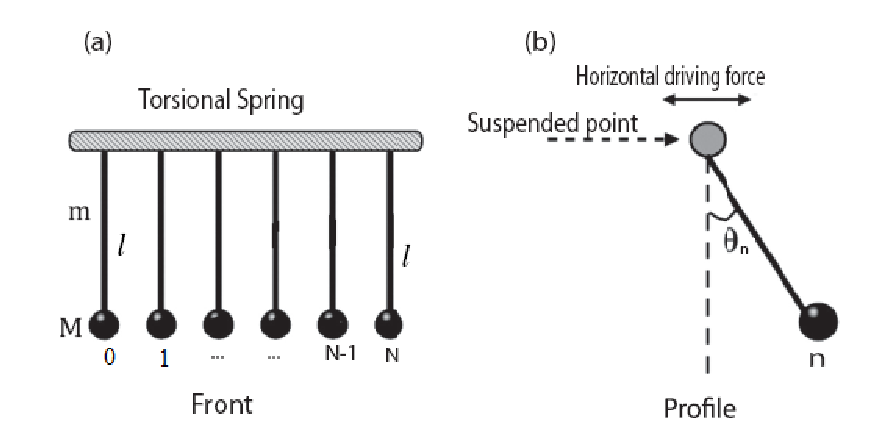}
\end{center}
 \caption{
Schematic representation of a pendulum chain connected by
torsional springs subjected to a horizontal driving force. (a)
front view; (b) profile view.  Reproduced with permission from
Ref.\,\cite{XuAlexanderKevrekidisPRE2014}, copyright by APS
2014.}\label{fig:model}
\end{figure}

The model under consideration consists of a pendulum chain
connected by torsional springs and subjected to a horizontal
driving force with frequency $\omega_{d}$ and amplitude $A$, as
illustrated  in Fig.\,\ref{fig:model}\, reproduced from
Ref.\,\cite{XuAlexanderKevrekidisPRE2014}. Each rigid rod of
length $l$  and mass $m$ supports the pendulum bob with mass $M$
at its end. The experimental system for the pendulum chain shown
in Fig.\,\ref{fig:model} had been proposed in
Ref.~\cite{BasuEnglish2008}. The Lagrangian for a chain of pendula
in absence of damping  can be written as follows
\cite{XuAlexanderKevrekidisPRE2014}:
\begin{equation} \label{equ:Lagrangian}
\begin{array}{l}
\mathcal{L} =\sum\limits_{n = 1}^N \{
\dfrac{1}{2}I\dot{\theta_{n}}^{2}+ \dfrac{1}{2}\left(
Ml+\dfrac{ml}{2}\right)g\cos(\theta_{n})~~~~~~~~~~~~~~~~~~~~~~~~~~~~~~\\~~~~~~~+\dfrac{1}{2}\left(
Ml+\dfrac{ml}{2}\right)\left[2A\omega_{d}\dot{\theta_{n}}\sin(\omega_{d}t_{1})\cos(\theta_{n})\right]\\
~~~~~~~-\dfrac{1}{4}\beta\left[\left(\theta_{n}-\theta_{n-1}\right)^{2}+\left(\theta_{n}-\theta_{n+1}\right)^{2}\right]\},
\end{array}
\end{equation}
with $I=Ml^{2}+\frac{1}{3}ml^{2}$ being the system moment of
inertia.
 The angle $\theta_{n}$
 measures the deviation from the vertical for the nth
pendulum, $\dot{\theta_{n}}$ is the corresponding angular speed,
 $g$ denotes the acceleration due to gravity, and $\beta$ is the linear
coupling coefficient between pendula due to a torsion spring.  The
equation of motion for the nth pendulum, derived using the
Euler-Lagrange's equation, is given by
\cite{CuevasPRL2009,BasuEnglish2008}:
\begin{equation} \label{equ:realmodel}
\begin{array}{l}
\ddot{\theta}_{n}-\frac{\beta}{I}
(\theta_{n+1}+\theta_{n-1}-2\theta_{n})+\omega_{0}^{2}\sin(\theta_{n})+~~~~~~~~~~\\~~~~~~~~~~~~~~~~~~~~~~~~~~~~~~f\omega_{d}^{2}\cos(\omega_{d}
t_{1})\cos(\theta_{n})=0,
\end{array}
\end{equation}
with $\omega_{0}^{2}=\dfrac{g}{I}(Ml+\frac{ml}{2})$, and
$f=\dfrac{\omega_{0}^{2}A}{g}$, being the dimensionless forcing
coefficient. Equation \eqref{equ:realmodel} can be further
simplified by scaling the time using the transformation.
$t_{1}=\dfrac{t}{\omega_{0}}$. In this way, the dimensionless form
of the equation of motion for the nth pendulum can be written as
follows:

\begin{equation} \label{equ1model}
\begin{array}{l}
\ddot{\theta}_{n}-c
(\theta_{n+1}+\theta_{n-1}-2\theta_{n})+\sin(\theta_{n})+~~~~~~~~~~\\~~~~~~~~~~~~~~~~~~~~~~~~~~~~~~f\omega_{1}^{2}\cos(\omega_{1}
t)\cos(\theta_{n})=0,
\end{array}
\end{equation}
where $c=\dfrac{\beta}{I}$
 is the dimensionless coupling parameter,
 and
$\omega_{1}=\dfrac{\omega_{d}}{\omega_{0}^{}}$ is the
dimensionless frequency for the periodic horizontal displacement
of the pivot.
 The linear dispersion relation for Eq.\,\eqref{equ1model} is given by
\begin{equation} \label{equ:dispersion}
\omega^{2}=1+2c(1-\cos(k))\, ,
\end{equation}
from which the linear band $1\leq\omega\leq\sqrt{1+4c}$ is obtained.

\subsection{\label{subsec:1}Supratransmission threshold }
\begin{figure}[ht]
\begin{center}
\includegraphics[width=3.0in]{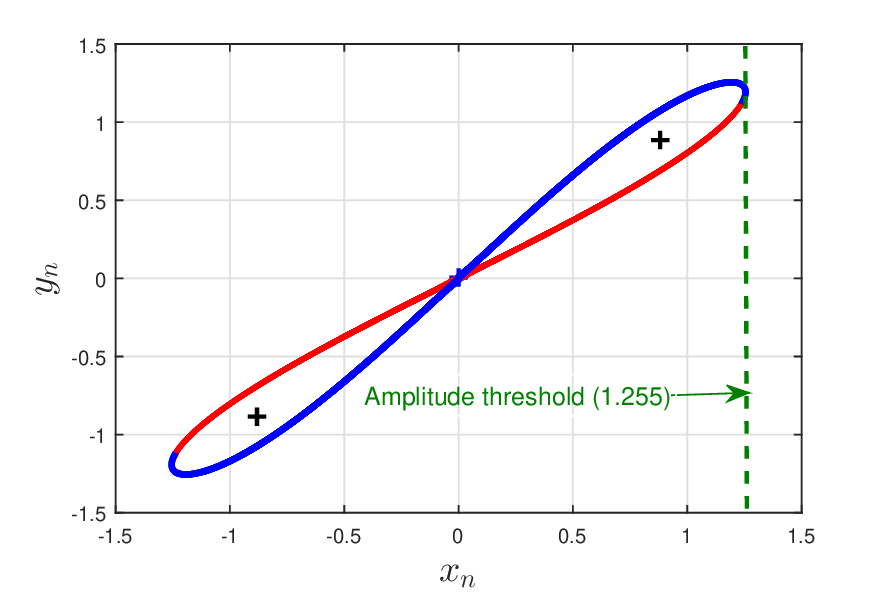}
\end{center}
\caption{Homoclinic orbit of  the 2D map \eqref{equ:2Dmap} for
$c=1$, and $\omega=0.95.$. The dashed green line corresponds to
the supratransmission threshold: $A_{thr}=1.255$.}
\label{fig:mapRogue}
\end{figure}

Here, we consider the model without the forcing coefficient, i.e.,
$f=0$, then Eq.\,\eqref{equ1model} becomes the discrete
Sine-Gordon equation. Expanding $\sin(\theta_{n})$ as  a Taylor
series up to the third order, the equations of motion become
\begin{equation} \label{equ:modeltaylor}
\ddot{\theta}_{n}-c
(\theta_{n+1}+\theta_{n-1}-2\theta_{n})+\theta_{n}-\frac{\theta_{n}^{3}}{6}=0.
\end{equation}
The time-periodic solution of the equation can be obtained by
proposing a harmonic solution in the form
$\theta_{n}=x_{n}\cos(\omega t)$ (see Refs.
\cite{KamdoumTogueuCSF22,PanagopoulosBountisSkokosJVA2004,RomeroRegaMecanica2015}).
The map corresponding to the stationary equation  can be written
as
\cite{KamdoumTogueuCSF22,PanagopoulosBountisSkokosJVA2004,RomeroRegaMecanica2015,
BountisCapelKollmannPLA2000,CarreteroChongPD06,PalmeroCarreteroChaptePRE2008,
CarreteroGonzalezChapter2009,TogueuPLA11,JohanssonKopidakisLepriAubryEPL2009}

\begin{equation}\label{equ:2Dmap}
{x_{n + 1}} = a{x_n} - bx_n^3 - {y_n}\,,\,\,\,\,\,\,\,\,\,{y_{n +
1}} = {x_n}.
\end{equation}
with $a  =  2 + \dfrac{1}{c}\left(1-\omega^{2} \right)$ and $b =
\dfrac{{1}}{{8c}}$. This map possesses three fixed points: $x_0=0$
and $x_\pm=\pm\sqrt{8(1-\omega^{2})}$.  The latter two exist
 only for $\omega<1$. This frequency band corresponds to the lower forbidden band, which is in agreement
with the band for which the supratransmission phenomenon has been
observed by  Geniet and L\'eon \cite{GenietLeonPRL02}.
 Fixed points are depicted by a cross in Fig.\,\ref{fig:mapRogue}.
 The stability of a fixed point is obtained by linearizing the map around the fixed point
  using the procedure described in Ref.\,\cite{TogueuPLA11}.
  A necessary condition for the existence of  manifolds is that
the fixed point must be a saddle point. A homoclinic orbit
corresponds to an orbit that connects, both in forward and
backward time, a saddle fixed point with itself
\cite{CarreteroGonzalezChapter2009}. Within the lower forbidden
band ($\omega<1$) only the fixed point $x_0=0$ is a saddle point.
Figure~\ref{fig:mapRogue} depicts the progression of the stable
(blue line) and unstable (red line) manifolds emanating from the
fixed point $x_0=0$. Stable and unstable manifolds emanating from
the fixed point $x_0=0$ intersect and form a homoclinic orbit,
which can be identified by the loop in Fig.~\ref{fig:mapRogue}.
The first-order connections of the loop correspond to the main
homoclinic orbit as can be clearly observed in
Fig.~\ref{fig:mapRogue}. The supratransmission threshold
corresponds to the value of the turning point of the main
homoclinic orbit \cite{TogueuGAPcncns17}. Its value can be
identified in this work by the dashed green line in
Fig.\,\ref{fig:mapRogue}.

\begin{figure}[ht]
\begin{center}
\includegraphics[width=3.0in]{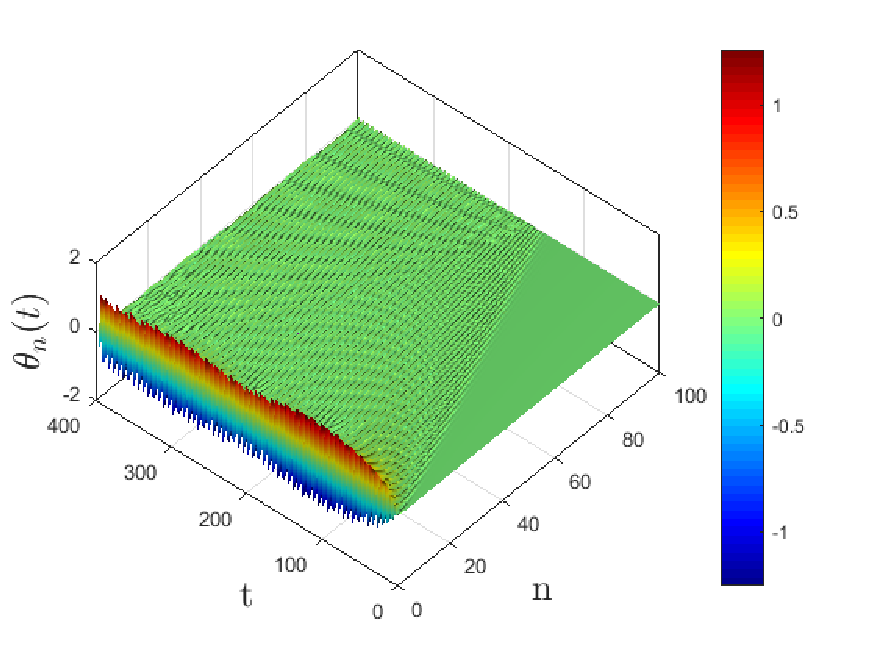}
\includegraphics[width=3.0in]{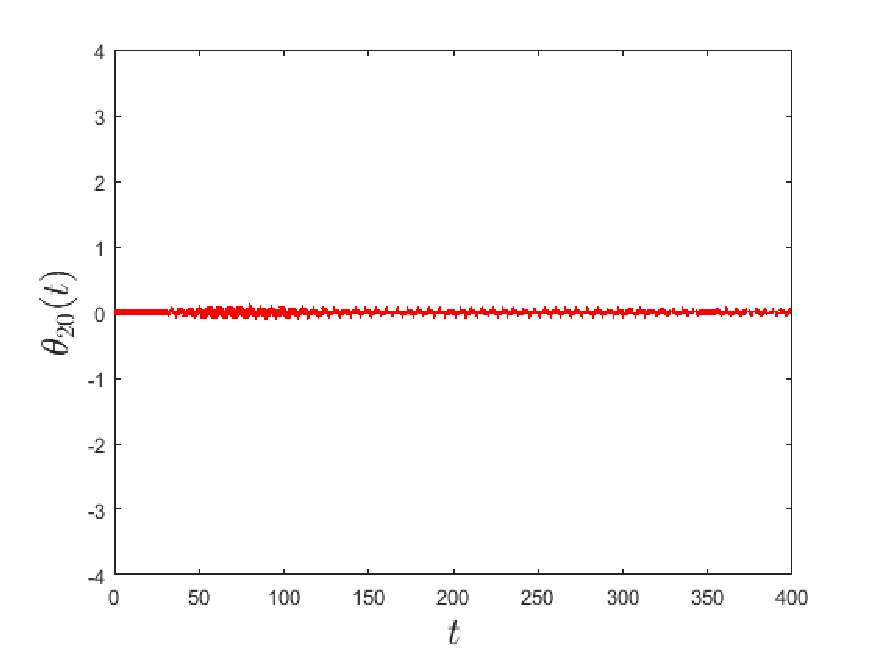}
\end{center}
\caption{Spatiotemporal evolution for the discrete equation
\eqref{equ1model} with boundary driving condition
\eqref{equ:driven}. The parameters are $f=0$,  $c=1$,
$\omega=0.95.$ and $A=1.254< A_{thr}$} \label{fig:suprabelow}
\end{figure}

\begin{figure}[ht]
\begin{center}
\includegraphics[width=2.6in]{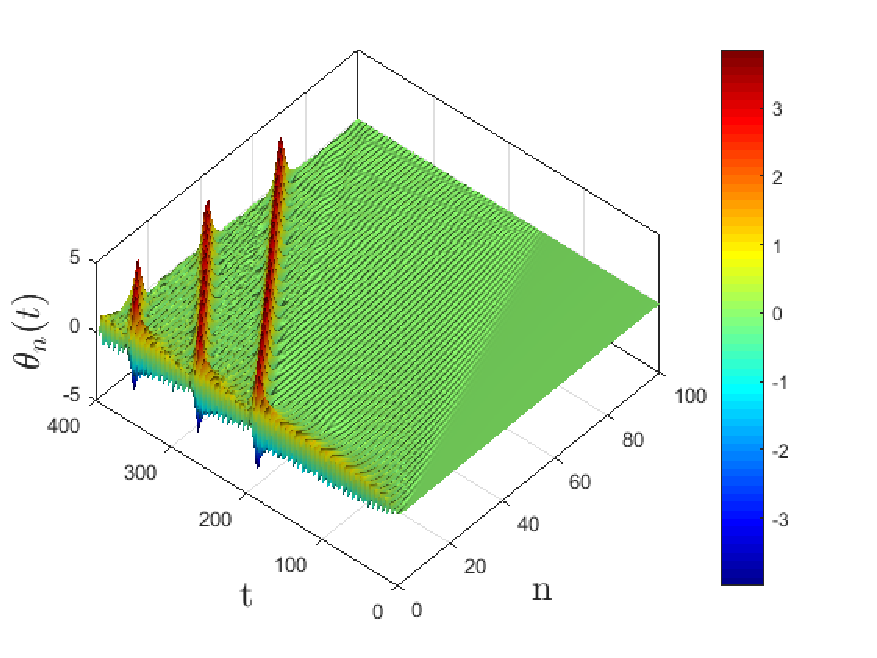}
\includegraphics[width=2.6in]{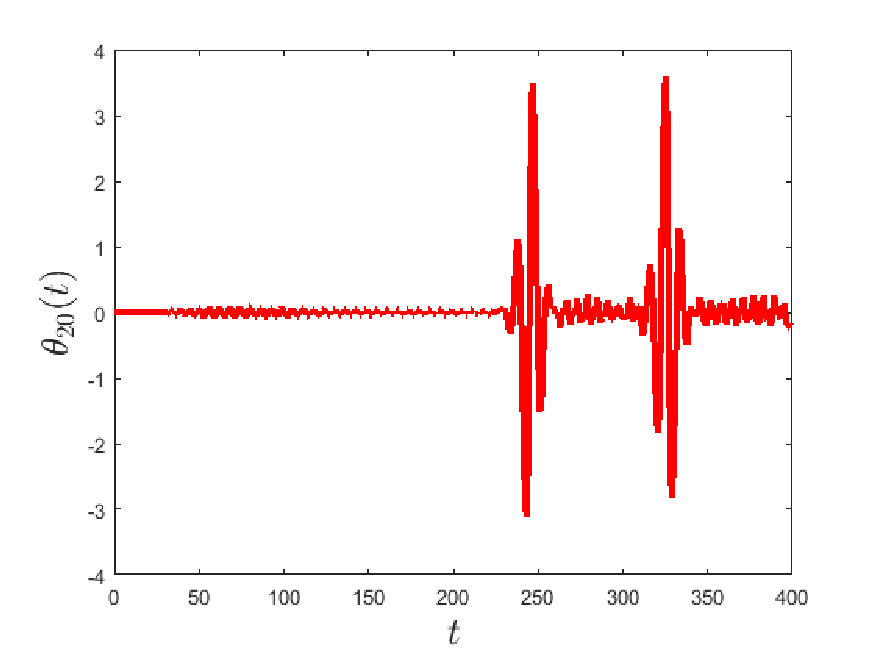}
\end{center}
\caption{Spatiotemporal evolution for the discrete equation
\eqref{equ1model} with boundary driving condition
\eqref{equ:driven}. The parameters are $f=0$,  $c=1$,
$\omega=0.95.$ and $A=1.256>A_{thr}$.} \label{fig:supraabove}
\end{figure}

\section{Numerical experiments}\label{sec:3}
 In this section, numerical studies are carried out on the discrete
equation \eqref{equ1model} with the ode45 solver of MatLab. This
solver uses variable time step to keep the desired relative and
absolute tolerance, which are set at $10^{-10}$. The left boundary
of the lattice depends on whether or not the lattice is driven.
The reflection at the right edge of the lattice will be avoided by
choosing a large value of N=101 and appropriate time for the full
integration of Eq.\,\eqref{equ1model}.

\subsection{\label{subsec:driven}Driven pendulum chain without shaking}

 The unshaken lattice ($f=0$) will be consider here and  the following harmonic
boundary condition is imposed to the chain for the full
integration of Eq.\,\eqref{equ1model}:
 \begin{equation}\label{equ:driven}
{\theta_0}(t) = {A}\cos ({\omega}t),
\end{equation}
where $A$ is the driving amplitude smoothly growing from the value
0 to $A$ and $\omega$ is the dimensionless frequency.
Figure~\ref{fig:suprabelow} depicts the behavior of the chain with
driving amplitude $A=1.254$ (slightly below the threshold 1.255)
and with the dimensionless  frequency $\omega=0.95$ within the
lower forbidden band, while Fig.\,\ref{fig:supraabove} shows a
train of gap solitons generated by driving the lattice with
amplitude A=1.256 (slightly above the threshold 1.255) and the
same frequency. The evanescent wave seen shown in
Fig.\,\ref{fig:suprabelow} and the energy flow shown in
Fig.\,\ref{fig:supraabove} confirm the agreement of the homoclinic
threshold shown in Fig\,\ref{fig:mapRogue} with the numerical
simulation.

\begin{figure}[ht]
\begin{center}
\includegraphics[width=2.6in]{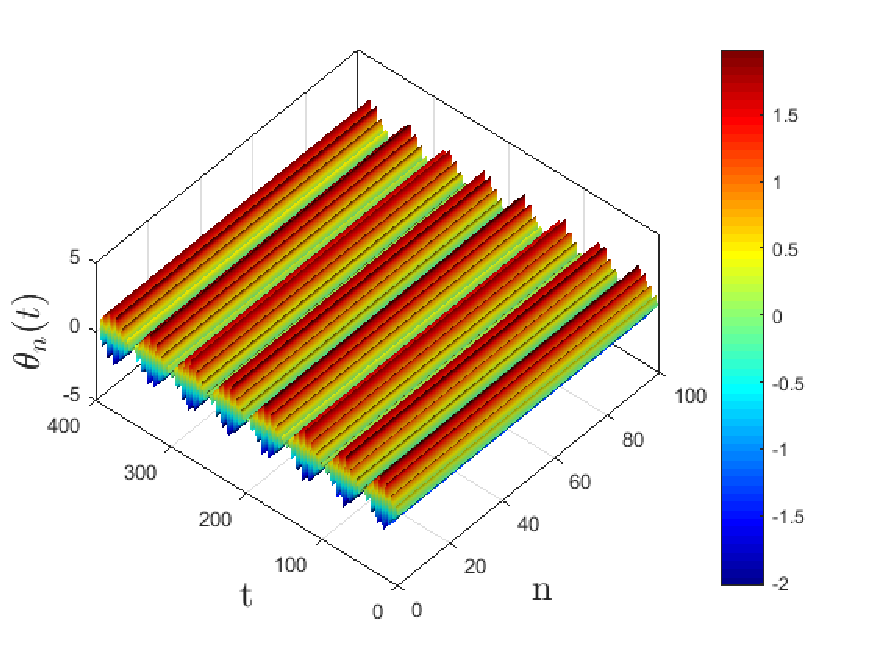}
\includegraphics[width=2.6in]{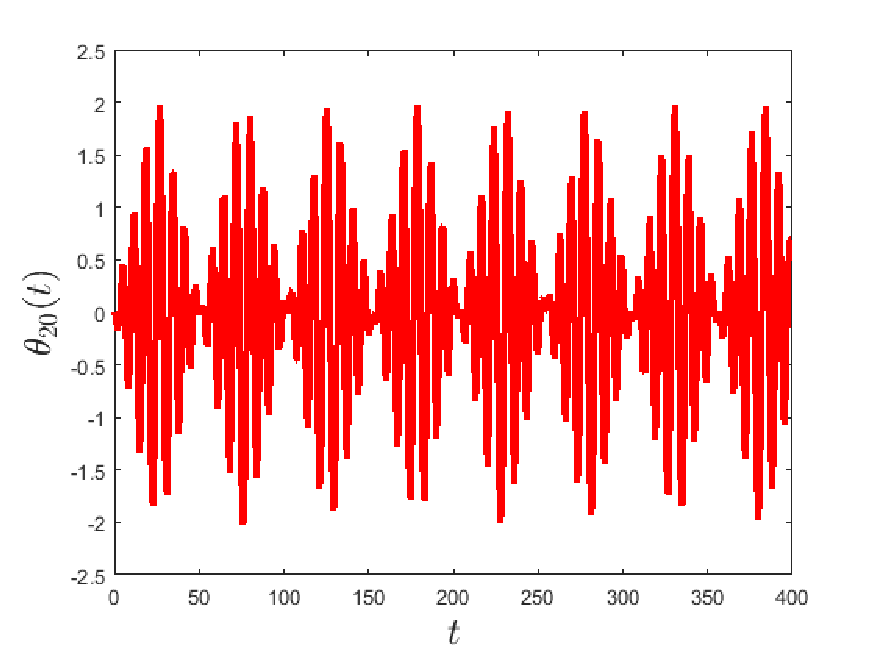}
\end{center}
\caption{Spatiotemporal evolution of the lattice submitted to
periodic horizontal shaking: $f=0.2$; $c=1$; $\omega_{1}=0.95$.}
\label{fig:Shakengap}
\end{figure}

\subsection{\label{subsec:2}Horizontally shaken pendulum chain}

The lattice is  subjected to a periodic horizontal displacement of
the pivot point with dimensionless frequency $\omega_{1}$ and
amplitude $f$. Figure~\ref{fig:Shakengap} depicts the
spatiotemporal evolution of the wave that results from the full
integration of Eq.\,\eqref{equ1model} with a shaking dimensionless
frequency in the lower forbidden band ($\omega_{1}=0.95$) and the
other parameters given by: $f=0.2$ and $c=1$. It is observed,
similarly to Ref.\,\cite{XuAlexanderKevrekidisPRE2014}, that the
lattice traps the gap soliton as and that, therefore, there is  no
transmission process. The same phenomenon is observed (not shown
here) in the upper forbidden band.

The trapping of the gap soliton here is an analogy with the light
being trapped into several neighboring waveguides obtained in
Ref.\,\cite{KhomerikiPRL04}, but it is done at different times.
The trapping here is different from the localization in space,
which is a consequence of the staggeringly driving force obtained
in Ref.\,\cite{CuberoCuevasPRL2009}. Another form of localization
can be obtained  using random fluctuations
\cite{SantisGuarcelloCSF2023}. By periodically driving the
lattice, the gap transmission is observed while the soliton is
trapped at a different time when the chain is subjected to a
periodic horizontal shaking. Below, the behavior of the chain will
be explored when it is subjected to being both simultaneously
driven and shaken.
\begin{figure}[t]
\begin{center}
\includegraphics[width=2.6in]{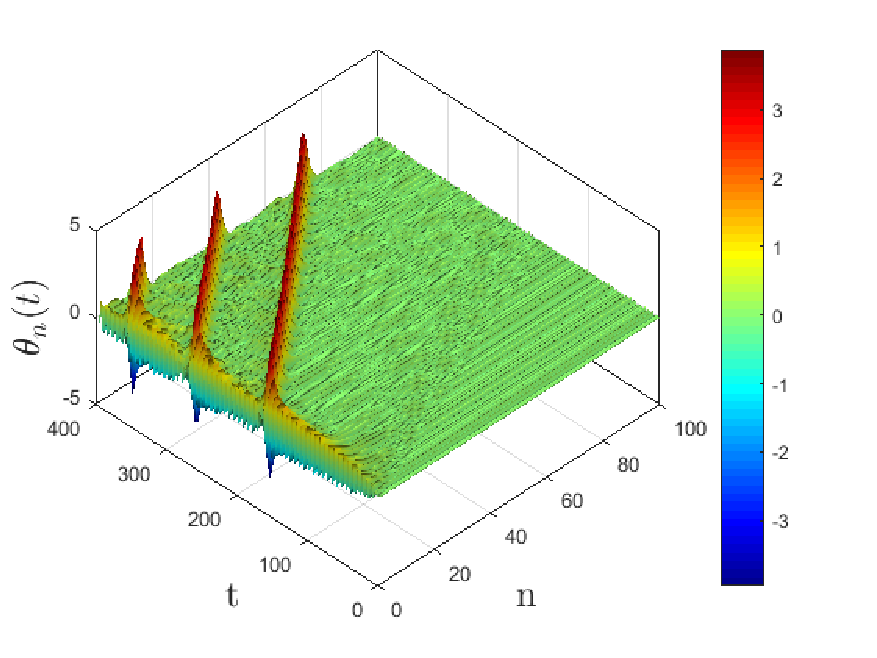}
\end{center}
\caption{Spatiotemporal evolution of the lattice with coupling
constant $c=1$ submitted simultaneously to a periodically driven edge
($\omega=0.95$, $A=1.22< A_{thr}$), and a periodic horizontal shaking
with frequency $\omega_{1}=1.22$  in the linear band. The forcing
coefficient is  $f=0.02$.} \label{fig:shakenPhonon}
\end{figure}

\subsection{\label{subsec:2}Simultaneously driven and shaken pendulum chain }
 Here we have two excitations: periodically driven edge and parametric
excitation. The periodically driven edge  frequency is taken in
the lower forbidden band ($\omega = 0.95$). The shaking frequency
$\omega_{1}$ will be taken  firstly within the linear phonon band
($1\leq\omega\leq\sqrt{1+4c}$) and secondly in the lower forbidden
band ($\omega<1$).

Figure~\ref{fig:shakenPhonon} shows the spatiotemporal evolution
of the chain with a driving frequency ($\omega=0.95$) within the
lower forbidden band and with driving amplitude ($A = 1.22$) below the
supratransmission threshold ($A_{thr}=1.255$). The shaking
frequency ($\omega_{1}=1.22$) is taken within the linear phonon
band. The generation of a train of solitons is observed, although
the driving amplitude is below the threshold. Therefore, the
presence of the shaking reduces the supratransmission threshold.
The same phenomenon is observed (not shown here) when the shaking
frequency is taken in the upper forbidden band.

\begin{figure}[ht]
\begin{center}
a)\includegraphics[width=2.6in]{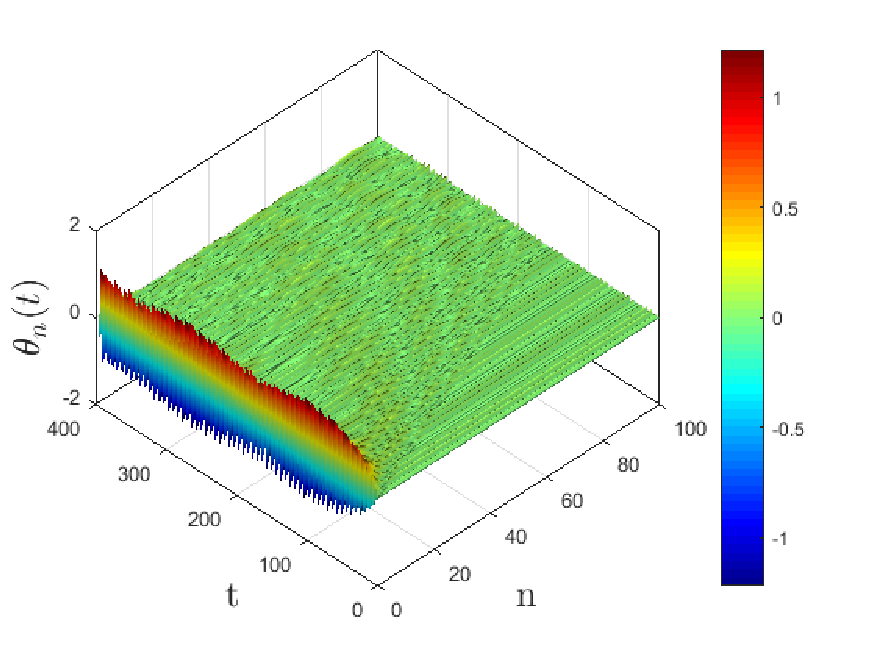}
b)\includegraphics[width=2.6in]{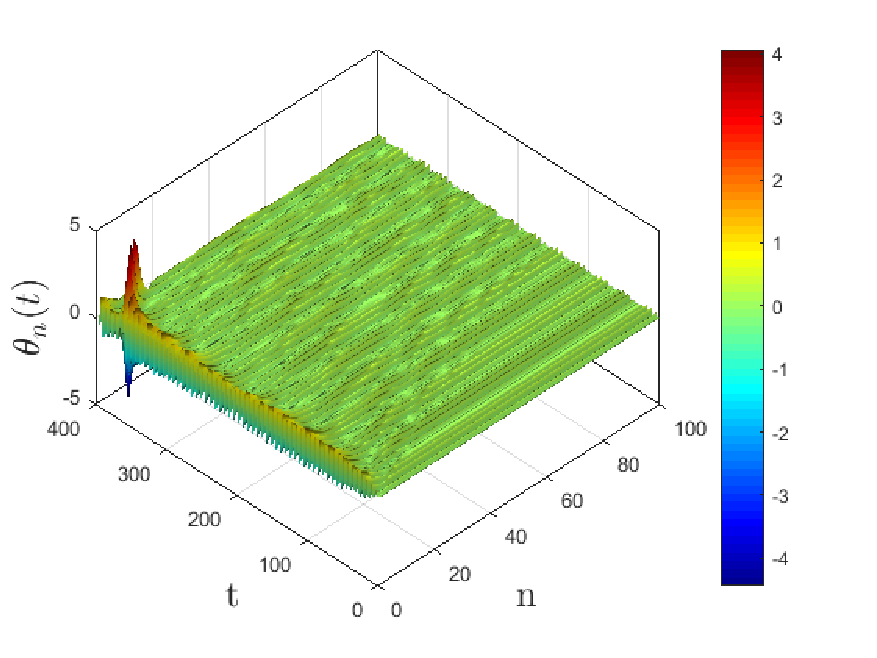}
\end{center}
\caption{Spatiotemporal evolution of the lattice with coupling
constant $c=1$ submitted simultaneously to a periodically driven edge
($\omega=0.95$, $A=1.256> A_{thr}$), and a periodic horizontal
shaking frequency $\omega_{1}=0.8$  in the lower forbidden band. Forcing coefficient: a)
$f=0.02$, b) $f=0.08$} \label{fig:driveshakenband}
\end{figure}

Let us now consider the case where the driving amplitude is $A =
1.256>A_{thr}$ and the frequency is $\omega=0.95$. Without the
shaking phenomenon ($f=0$), numerical integration of
Eq.\,\eqref{equ1model} with these parameters generates band-gap
transmission as can be seen in Fig.\,\ref{fig:supraabove}.
Figure~\ref{fig:driveshakenband}~(a) shows that the presence of
shaking with frequency $\omega_{1}=0.8$, within the lower
forbidden band, destroys the supratransmission phenomenon. This is
contrary to the case with the shaking frequency within the phonon
band.  When the forcing  coefficient $f$ increases to $f=0.08$,
the nonlinear bandgap transmission occurs as can be seen in
Fig.\,\ref{fig:driveshakenband}~(b). The presence of the shaking
phenomenon here increases the supratransmission threshold.

\begin{figure}[ht]
\begin{center}
a)\includegraphics[width=2.8in]{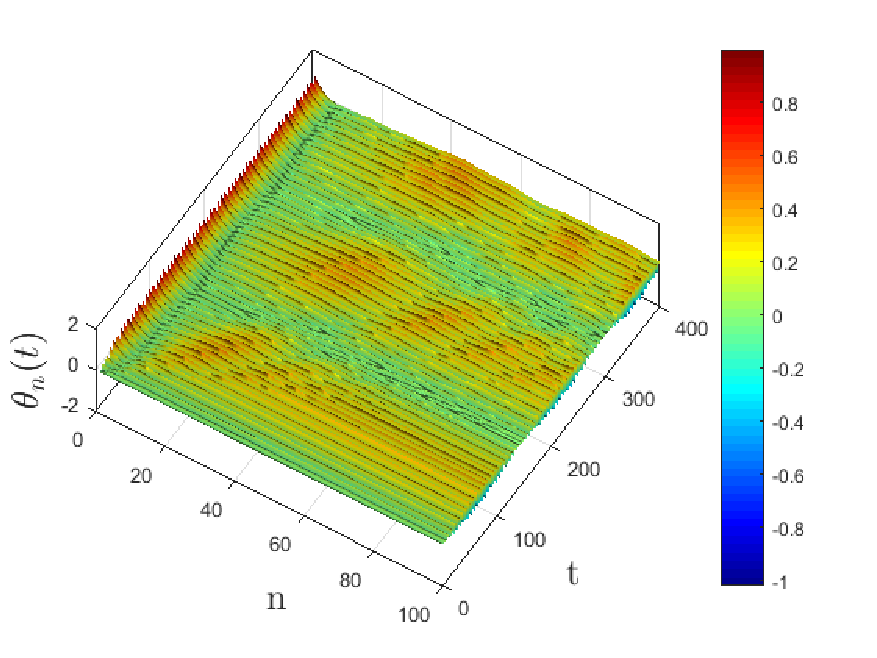}
b)\includegraphics[width=2.8in]{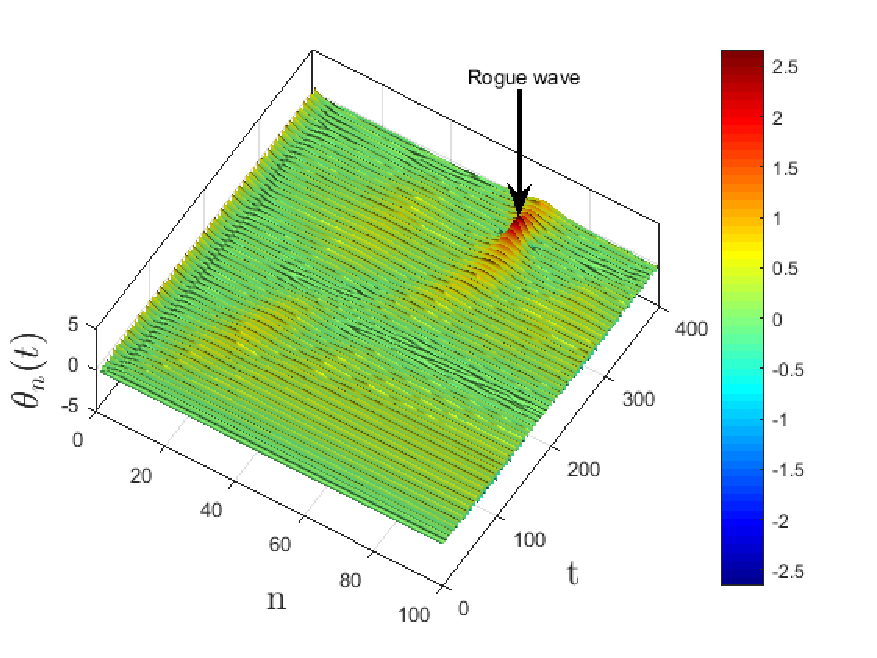}
c)\includegraphics[width=2.8in]{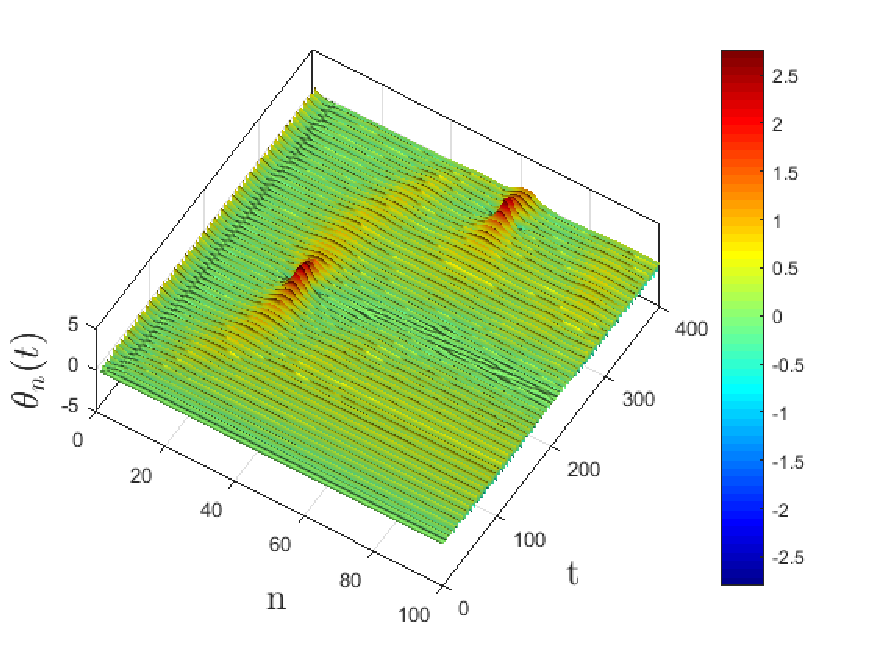}
d)\includegraphics[width=2.8in]{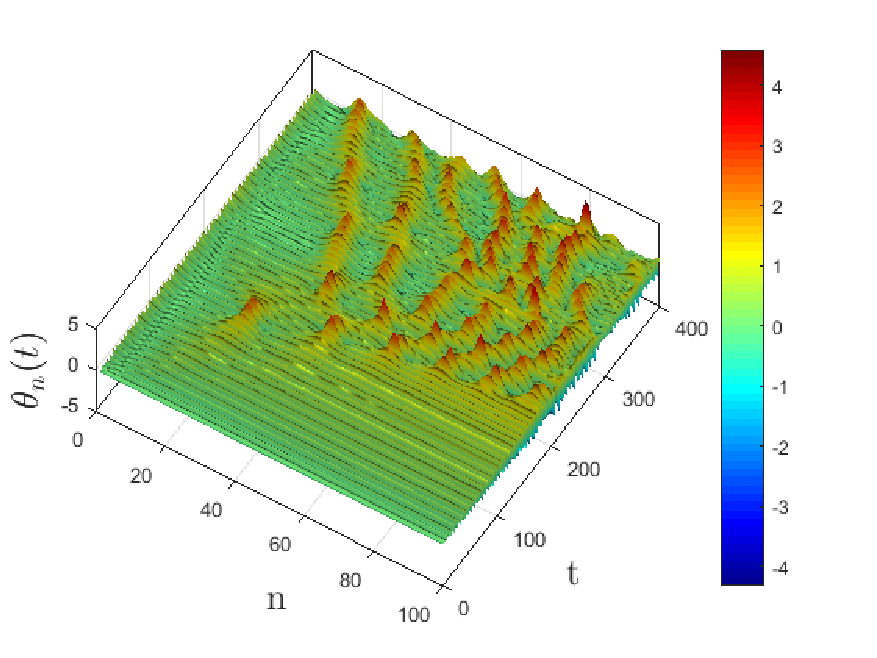}
\end{center}
\caption{Spatiotemporal evolution of the lattice with coupling
constant $c=1$ submitted simultaneously to a periodically driven
edge ($\omega=0.95$, $A=1.256$) and a periodic horizontal shaking
($\omega_{1}=\omega=0.95$) with the forcing coefficient $f$
increasing from left to right and from top to bottom: a) $f=0.02$,
b) $f=0.026$, c) $f=0.027$, d) $f=0.03$.} \label{fig:Simultaneous}
\end{figure}

In Fig.\,\ref{fig:Simultaneous}, we depict a progression of the
spatiotemporal evolution of the driven lattice  in the presence of
the shaking phenomenon for different forcing coefficients $f$. The
shaking frequency is  equal to the driving frequency $\omega$. For
small $f$ ($f = 0.02$), the band-gap transmission disappears,
although the driving frequency is within the gap and the amplitude
is above the supratransmission threshold, as can be seen in
Fig.\,\ref{fig:Simultaneous}~(a). The disappearance of the
supratransmission phenomenon means that the threshold increases in
the presence of shaking with frequency within the lower forbidden
band.

For  $f = 0.026$ (See Fig.\,\ref{fig:Simultaneous}~(b)),  an
unpredictable and unexpectedly localized wave appears and
disappears without a trace.  The phenomenon is similar to that
obtained by Akhmediev et al.~\cite{AkhmedievPLA2009} in a
continuous lattice. The discrete rogue wave is produced as a
result of  a simultaneously driven and shaken pendulum. For
$f=0.027$ (Fig.\,\ref{fig:Simultaneous}~(c)), the number of
unpredictable localized waves increases and the spatiotemporal
dynamic of the lattice is similar to the second-order discrete
rogue wave found analytically in
Ref.\,\cite{YongMalomedCHAOS2016}. For a slightly larger value of
the forcing coefficient (see Fig.\,\ref{fig:Simultaneous}~(d)),
 several localized waves appear  and
the configuration is similar to that obtained in  Ref.\,\cite{SotoCrespoPRL2016}.

A discrete rogue wave is generated here with a plane wave as the
initial condition. To the best of our knowledge, this is the first
time that this phenomenon is observed. It is worth pointing out
that presently it is not possible to discard  that a rogue wave
can also be generated in the just driven, not shaken case,
although it has not yet been observed.

\subsection{\label{subsec:fft} Spectral analysis}

\begin{figure}[t]
\begin{center}
\includegraphics[width=3.1in]{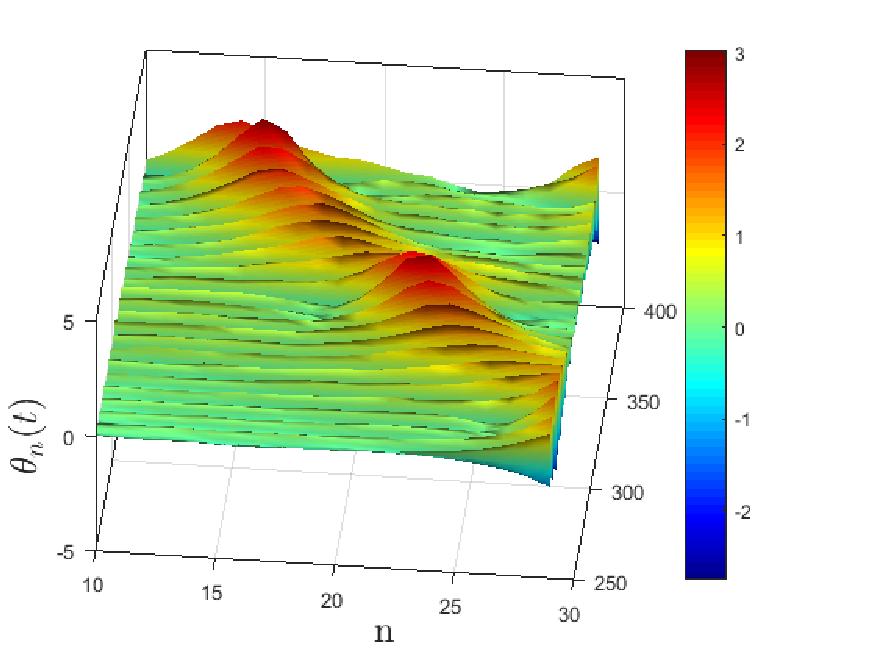}\\
\includegraphics[width=2.6in]{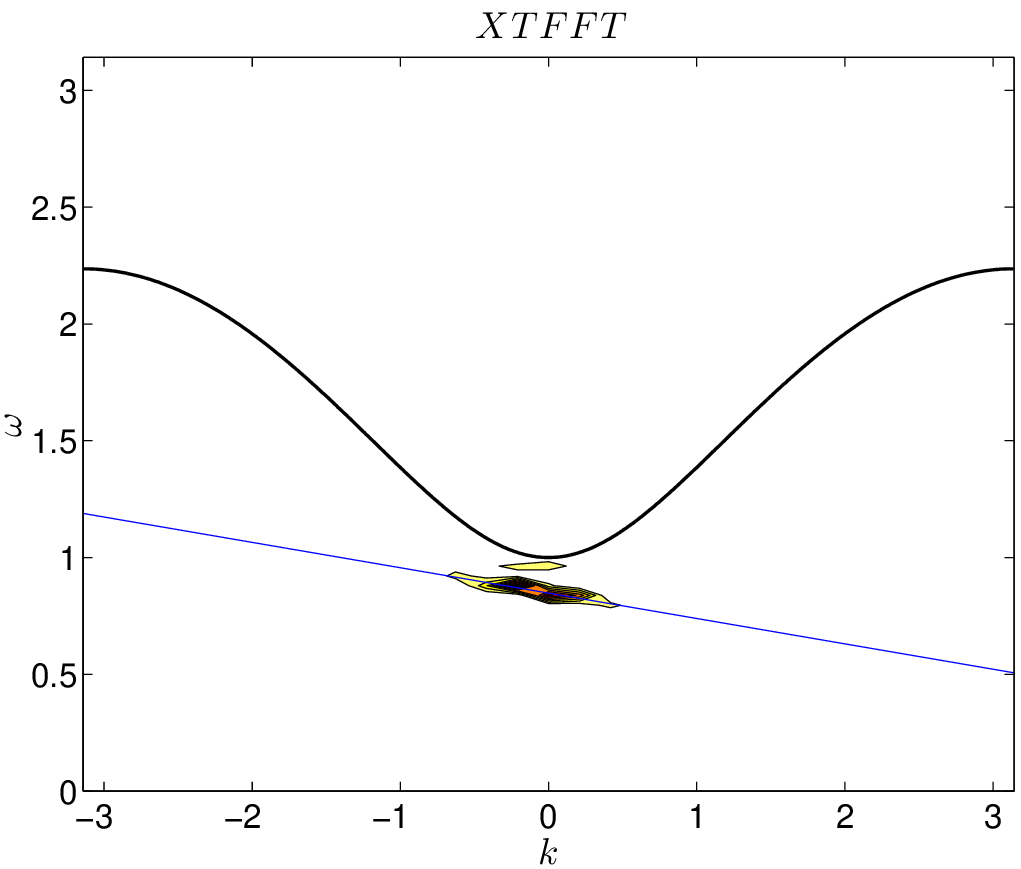}
\end{center}
\caption{({\bf Top}) Detail of the rogue wave observed in
Fig.\,\ref{fig:Simultaneous}~(d). ({\bf Bottom})
Frequency-momentum plot obtained using the two dimensional fast
Fourier transform in space and time (XTFFT) for the first 30 sites
and for a time between 250 and 400. The dispersion relation is
shown for reference. The narrow band with constant slope
corresponds to a breather with velocity $V_b=\dfrac{\partial
\omega}{\partial k}$. The driving and shaking frequency appears as
a faint short horizontal line just below the phonon band. See text
and Refs.\,\cite{archilla2019-23}. } \label{fig__XTFFTrogue}
\end{figure}

We have performed the spectral analysis of rogue waves following
Refs.\,\cite{archilla2019-23}. In particular we present here the
analysis of the case presented in Fig.\,\ref{fig:Simultaneous}~(d)
for $f=0.03$, where we can observe several rogue waves. As rogue
waves are limited in time and space the precision of the two
dimensional fast Fourier transform in space and time (XTFFT) is
small as it depends on the time interval and sublattice size,
which cannot be enlarged, but it provides valuable information
nonetheless. For example, we concentrate in the rogue wave
appearing in the first 30 particles for the time between 250 and
400, shown in Fig.\,\ref{fig__XTFFTrogue}~(a). The dispersion
relation and  the XTFFT are represented in
Fig.\,\ref{fig__XTFFTrogue}~(b). The breather band can be seen
slightly below the forcing frequency near $k=0$. The corresponding
breather line is also represented, its slope being the rogue wave
velocity $V_b\simeq 0.11$ which can also be obtained from the
$x(t)$ curve in Fig.\,\ref{fig__XTFFTrogue}~(a). The breather band
value for $k=0$ provides the frequency in the moving frame
$\Omega_b\simeq 0.85$. As this frequency is not zero, the rogue
wave is breather--like and comparing $\Omega_b$ with the velocity
frequency $\omega_V=2\pi V_b$ (the frequency at which a rogue wave
encounters different sites), we obtain that
$\Omega_b/\omega_V\simeq 5/4$. This means that the rogue wave
performs approximately 5 oscillations while moving 4 sites. As the
central wave vector of the breather band is near $k=0$, the
different pendula have a small phase difference, meaning that the
breather profile is bell shaped. Therefore, it is also confirmed
that in this system rogue waves are below the phonon band. Similar
analysis can be performed for other rogue waves with qualitatively
similar results.

\section{CONCLUSIONS}\label{sec:4}
In this work, we have studied the spatiotemporal behavior of the
discrete pendulum chain, firstly excited by a harmonic wave, then
by a parametric excitation, and finally by simultaneous driving and
shaking excitations. The one-edge-driven lattice produces the
well-known supratransmission phenomenon in agreement with the
homoclinic threshold. Localization in space of the wave is
obtained after shaking the lattice. The threshold of the
supratransmission is reduced when the lattice is simultaneously
driven with a frequency within the forbidden band and shaken with
a frequency within the phonon band. The threshold  value increases when it is
shaken with a frequency in the lower forbidden band.  Discrete
rogue waves are obtained after simultaneously driving and shaking
the lattice with the same frequency. The experimental device for
the shaken pendulum has been realized in order to derive discrete
breathers at Dickinson College\,\cite{EnglishPalmero}. We hope that the
result presented in this letter will allow the modification of the experimental setup in order to obtain
discrete rogue waves.

\onecolumngrid

\twocolumngrid

\section*{Acknowledgments}
Alain Bertrand Togueu Motcheyo wishes  to express its deepest
gratitude to Prof. Masayuki Kimura and  to all the organizers of
the "International Symposium on Nonlinear Theory and Its
Applications" (NOLTA 2022), December 12-15, 2022, for the
opportunity they provided to him for presenting a part of this work.
JFRA acknowledges the Universities of Osaka and Latvia for hospitality.

\section*{Funding}
MK acknowledges support from grants  JSPS Kakenhi (C) No. 21K03935.
YD  acknowledges the support from grant JSPS Kakenhi (C) No. 19K03654.
JFRA  thanks  projects  MICINN PID2019-109175GB-C22 and PID2022-138321NB-C22, and several travel grants
from the VII PPITUS-2023 of the University of Sevilla.

\end{document}